\documentclass[a4paper,12pt]{article}
\usepackage[utf8]{inputenc}
\usepackage[english]{babel}
\usepackage{authblk}
\usepackage{graphicx}
\usepackage{mathptmx}
\usepackage[singlespacing]{setspace}
\usepackage[headheight=1in,margin=1in]{geometry}
\usepackage{fancyhdr}
\usepackage{lipsum}

\graphicspath{{images/}}

\title{Evaluating internal and external dissonance of belief dynamics in social systems}

\author[1]{Joshua T. S. Hewson}
\author[2]{Ke Fang}
\affil[1]{Department of Cognitive, Linguistic, and Psychological Sciences, Brown University}
\affil[2]{Collective Cognition Lab, New York University}

\date{}

\begin{document}

\maketitle
\thispagestyle{fancy}

\begin{center}
\textit{Keywords: Belief dynamics, Belief network, Social network, Agent-based model}
\newline
\end{center}


\section*{Extended Abstract}

Belief dynamics are central to human society. Individuals depend on accurate beliefs to engage in appropriate behaviors (Hemmer et al., 2022; Yon et al., 2020), such as deciding which college or which job position to apply for. Additionally, beliefs are inherently social — successful coordination and cooperation are founded on shared beliefs, and people form groups based on deep-seated beliefs (Van Prooijen \& Douglas, 2018; White et al., 2021). However, beliefs were traditionally often studied in isolation; psychological studies typically focused on a single belief of independent individuals. However, this approach is increasingly challenged. On the one hand, the network perspective of individual belief systems suggests that beliefs should be considered interconnected systems -- a change in one belief is constrained by surrounding beliefs and leads to changes in other beliefs (Brandt, 2022; Dalege et al., 2016; Powell et al., 2023). On the other hand, a social network approach suggests groups can be viewed as a social network of beliefs held by individuals (Lüders et al., 2024; Sîrbu et al., 2017). Therefore, it is proposed that belief dynamics should increasingly be understood as a "network within a network," where the within-person belief network is embedded within between-person social networks (Galesic et al., 2021; Vlasceanu et al., 2023).

A fundamental force driving belief dynamics is the human endeavor to resolve inconsistencies within both individual and social networks (Dalege et al., 2016; Galesic et al., 2021). This concept is supported by classic psychological theories, such as cognitive dissonance (Festinger, 1957) and Heider's balance theory (Heider, 1958). However, the nature of the trade-offs between internal dissonance and social imbalance remains unclear. Existing models often presume that individuals have a fixed preference for resolving internal dissonance over addressing social imbalance (Galesic et al., 2021), but the validity of this assumption is debatable. In reality, individuals may navigate these inconsistencies in a more dynamic and context-dependent manner. Humans are capable of synthesizing advice and personal experience, typically leaning towards more reliable sources (Biele et al., 2009; Vélez \& Gweon, 2019). For instance, if an individual encounters conflicting beliefs about a particular issue but the social consensus on that issue is clear, they may choose to align with the social norm. Conversely, if the social consensus is ambiguous, but their personal beliefs are consistent, it may be more rational to trust their own judgment. This concept mirrors the statistical principle of inverse-variance weighting (IVW). Under the assumption of normal distribution, IVW provides the maximum likelihood estimate of the true value. Furthermore, considering the growing evidence that humans are approximate Bayesian learners, the IVW estimate can also be interpreted as the posterior distribution for the true value, assuming normally distributed observations and a non-informative prior.

To test the effect of IVW of dissonances on belief dynamics, we proposed an extension of the model from Galesic et al.(2021). One of the primary assumptions held by this model is that people want to reduce dissonance across beliefs, both internally and across their social network. When considering a specific focal belief within a network, people can reduce dissonance by either updating that belief or by updating their social network, potentially cutting ties with people who disagree. When considering the total dissonance, one important factor is the extent to which people weigh their own internal dissonance relative to the dissonance of their neighbors in the social network. Galesic et al. (2021) defined this as a constant (denoted as w). When w is higher, people prefer their beliefs to be aligned with the beliefs of their friends, even if internal dissonance remains. Dissonance within a network is defined as the variance between each belief and the beliefs neighboring it. For each belief, an individual’s views can be modeled by a probability distribution showing the extent to which they disagree or agree. When there is a large amount of variation in the extent to which people agree or disagree with related beliefs within a belief network, belief dissonance for an individual is high. When there is variation in the extent to which people hold certain beliefs across a social network, social dissonance is high.

Our extension of the model redefines w, for each individual, as a function of the internal and social belief certainty. We modeled certainty as the variance across an individual’s internal beliefs and the variance across one’s social contacts. The lower the variance across beliefs, the higher the certainty. When an individual’s beliefs are all aligned with each other, the variance across these beliefs approaches 0. When there is variation across an individual’s beliefs, the variance increases. As opposed to the original model, in our modified version, we change the focal belief at every iteration. This modification results in consistently decreasing dissonance so that it reaches an equilibrium state where the dissonance is minimized.

The results suggest two different patterns of belief convergence, with w either converging to zero or one. When w converges to zero, meaning that the social dissonance is ignored across all individuals, the resulting state is seen in the left heatmap in Figure 1, where all beliefs become internally aligned. Despite having opposing beliefs of neighbors in the social network, people’s internal beliefs converge to a particular belief level which is the same across each belief in the belief network. This can be thought of as having ideologically consistent but socially disagreeable beliefs. On the other hand, if w converges to one, meaning that the agents only weigh social dissonance, the resulting belief states is presented in the right heatmap in Figure 1, where all beliefs become socially aligned. In this case, since agents disregard any internal dissonance, each belief aligns between the social neighbors although the beliefs of an individual varies for each distinct belief in the belief network. This can be thought of as having agreeable but inconsistent beliefs. Typically there is a period where w values fluctuate before converging on a value (Figure 2).

We believe this convergence occurs because of a positive feedback loop between increasing confidence and decreasing dissonance in one of the two networks, resulting in a momentary and unstable competition between the two networks, where one network quickly becomes ignored. Our preliminary analysis suggests the size and connectivity of the two networks determine whether the model collapses in favor of the internal network (w=0) - resulting in internal alignment of beliefs - or the social network (w=1) - resulting in the alignment of beliefs socially.

Overall, we extended Galesic et al.'s model of belief dynamics by allowing for individual's to weigh up internal and social dissonance in the context of their certainty, better modeling a psychologically realistic approximate bayesian learner. Interestingly we see that when individuals can determine how they weigh up their own beliefs with their neighbors, their internal network and their social network enter into competition for attention. However, the tendency for the model to converge, and to only do so at the extremes suggests there is a lack of negative feedback built into the model. Looking forward, we hope to address this issue so that our model better reflects the ongoing dynamics of beliefs seen in society.

\begin{figure}[htp]
    \centering
    \small
    \includegraphics[width=0.4\textwidth]{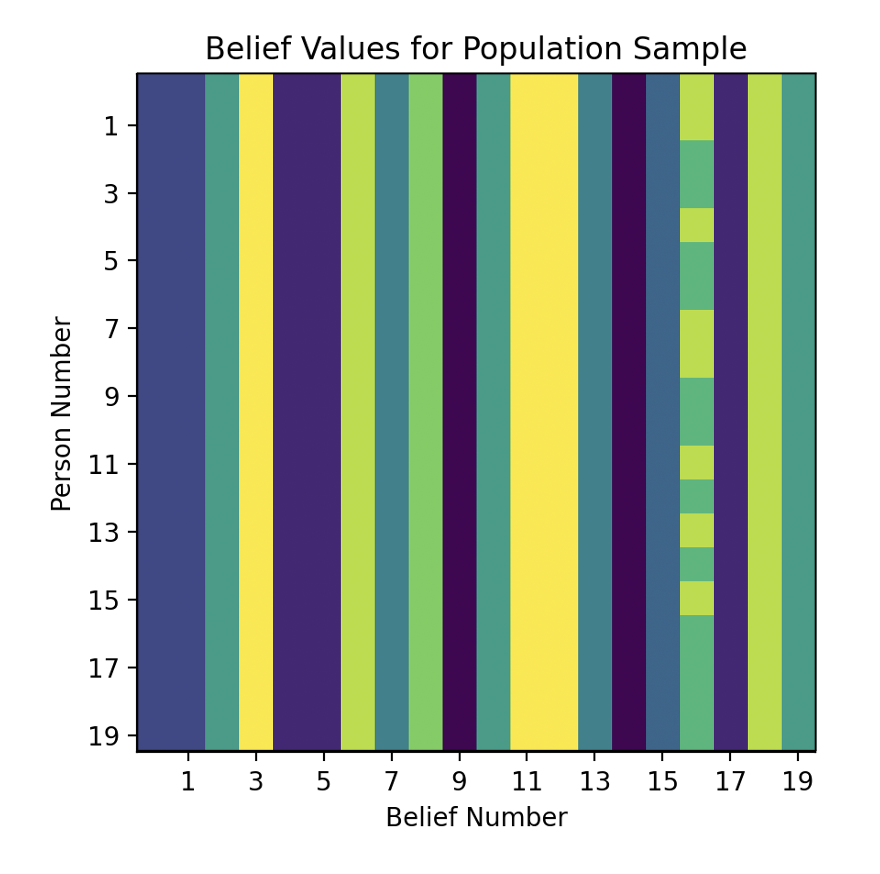}
    \includegraphics[width=0.4\textwidth]{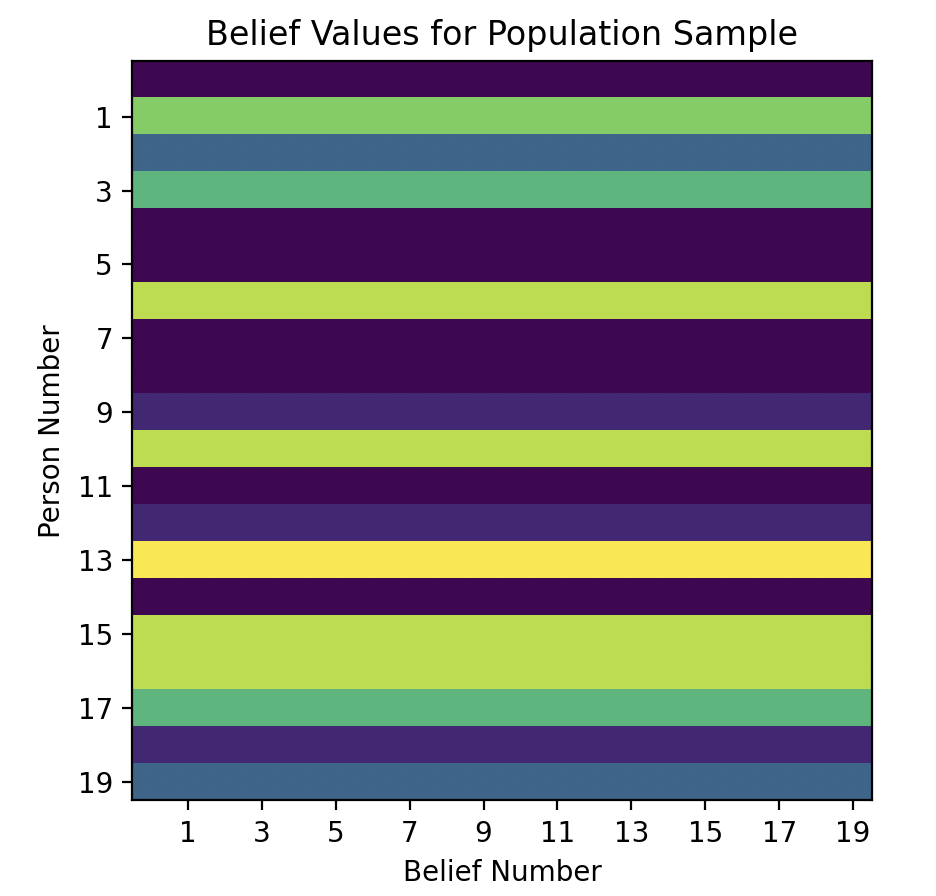}
    \caption{Belief levels at the end of the simulation for 20 agents i) for w = 0 on the left ii) for w = 1 on the right 
}
    \label{fig:image}
\end{figure}

\begin{figure}[htp]
    \centering
    \includegraphics[width=0.6\textwidth]{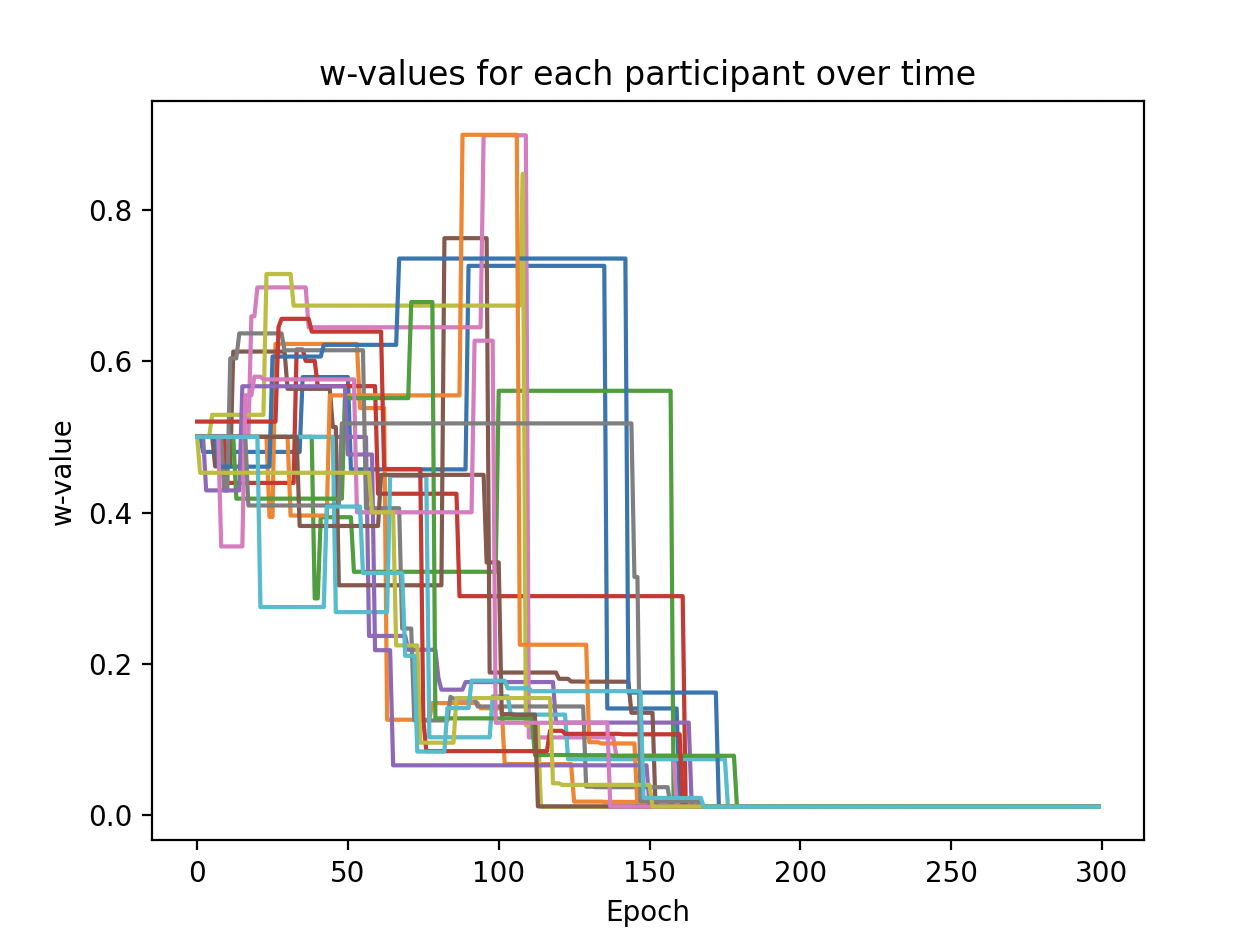}
    \caption{w values for each agent throughout the simulation. Here we see an example where the w-values converge to zero, producing an array of belief values as seen in Figure 1 (ii)}
    \label{fig:image}
\end{figure}

\section*{References}

Biele, G., Rieskamp, J., \& Gonzalez, R. (2009). Computational Models for the Combination of Advice and Individual Learning. Cognitive Science, 33(2), 206–242. https://doi.org/10.1111/j.1551-6709.2009.01010.x

Brandt, M. J. (2022). Measuring the belief system of a person. Journal of Personality and Social Psychology, 123(4), 830–853. https://doi.org/10.1037/pspp0000416

Dalege, J., Borsboom, D., van Harreveld, F., van den Berg, H., Conner, M., \& van der Maas, H. L. J. (2016). Toward a formalized account of attitudes: The Causal Attitude Network (CAN) model. Psychological Review, 123(1), 2–22. https://doi.org/10.1037/a0039802

Galesic, M., Olsson, H., Dalege, J., van der Does, T., \& Stein, D. L. (2021). Integrating social and cognitive aspects of belief dynamics: Towards a unifying framework. Journal of The Royal Society Interface, 18(176), 20200857. https://doi.org/10.1098/rsif.2020.0857

Hemmer, P., Musolino, J., \& Sommer, J. (2022). Introduction: Toward a Cognitive Science of Belief. In P. Hemmer, J. Musolino, \& J. Sommer (Eds.), The Cognitive Science of Belief: A Multidisciplinary Approach (pp. 1–28). Cambridge University Press; Cambridge Core. https://doi.org/10.1017/9781009001021.001

Lüders, A., Carpentras, D., \& Quayle, M. (2024). Attitude networks as intergroup realities: Using network‐modelling to research attitude‐identity relationships in polarized political contexts. British Journal of Social Psychology, 63(1), 37–51. https://doi.org/10.1111/bjso.12665

Powell, D., Weisman, K., \& Markman, E. M. (2023). Modeling and leveraging intuitive theories to improve vaccine attitudes. Journal of Experimental Psychology. General, 152(5), 1379–1395. https://doi.org/10.1037/xge0001324

Sîrbu, A., Loreto, V., Servedio, V. D. P., \& Tria, F. (2017). Opinion dynamics: Models, extensions and external effects (pp. 363–401). https://doi.org/10.1007/978-3-319-25658-0\_17
Van Prooijen, J.-W., \& Douglas, K. M. (2018). Belief in conspiracy theories: Basic principles of an emerging research domain. European Journal of Social Psychology, 48(7), 897–908. https://doi.org/10.1002/ejsp.2530

Vélez, N., \& Gweon, H. (2019). Integrating Incomplete Information With Imperfect Advice. Topics in Cognitive Science, 11(2), 299–315. https://doi.org/10.1111/tops.12388

Vlasceanu, M., Dyckovsky, A. M., \& Coman, A. (2023). A Network Approach to Investigate the Dynamics of Individual and Collective Beliefs: Advances and Applications of the BENDING Model. Perspectives on Psychological Science, 17456916231185776.\\ https://doi.org/10.1177/17456916231185776

White, C. J., Baimel, A., \& Norenzayan, A. (2021). How cultural learning and cognitive biases shape religious beliefs. Current Opinion in Psychology, 40, 34–39.\\ https://doi.org/10.1016/j.copsyc.2020.07.033

Yon, D., Heyes, C., \& Press, C. (2020). Beliefs and desires in the predictive brain. Nature Communications, 11(1), 4404. https://doi.org/10.1038/s41467-020-18332-9

\end{document}